\documentclass[doublecol]{epl2} 

\usepackage{amsfonts,amsmath}

\usepackage{epsfig}
\usepackage{color}
\usepackage{bm}

\usepackage{braket}

\usepackage{comment}

\usepackage{graphicx}
\usepackage{amsthm}

\newcommand{\eq}[1]{\begin{equation} #1 \end{equation}}
\newcommand{\eqa}[2]{\begin{equation} #1 \label{#2} \end{equation}}
\newcommand{\balign}[1]{\begin{align} #1 \end{align}}
\newcommand{\bcases}[1]{\begin{cases} #1 \end{cases}}

\newcommand{\mx}[1]{\begin{pmatrix}#1 \end{pmatrix}}

\newcommand{\todayd}{\the\year/\the\month/\the\day}

\newcommand{\bib}{\bibitem}

\newcommand{\lmd}{\lambda}

\newcommand{\lb}{\label}

\newcommand{\Tr}{\mathrm{Tr}}

\newcommand{\bel}{\begin{easylist}}
\newcommand{\eel}{\end{easylist}}

\newcommand{\eref}[1]{Eq.~\eqref{#1}}
\newcommand{\tref}[1]{Theorem.~\ref{t:#1}}

\newcommand{\lref}[1]{Lemma.~\ref{t:#1}}
\newcommand{\pref}[1]{Proposition.~\ref{t:#1}}

\def \({\left(}
\def \){\right)}
\def \[{\left[}
\def \]{\right]}

\newcommand{\abs}[1]{\left|#1\right|}

\newcommand{\sumtwo}[2]%
{\mathop{\sum_{#1}}_{#2}}
\newcommand{\sumthree}[3]%
{\mathop{\mathop{\sum_{#1}}_{#2}}_{#3}}
\newcommand{\sumfour}[4]%
{\mathop{\mathop{\mathop{\sum_{#1}}_{#2}}_{#3}}_{#4}} 
\newcommand{\prodtwo}[2]%
{\mathop{\prod_{#1}}_{#2}}
\newcommand{\mintwo}[2]%
{\mathop{\min_{#1}}_{#2}}
\newcommand{\maxtwo}[2]%
{\mathop{\max_{#1}}_{#2}}
\newcommand{\maxthree}[3]%
{\mathop{\mathop{\max_{#1}}_{#2}}_{#3}}
\newcommand{\limtwo}[2]%
{\mathop{\lim_{#1}}_{#2}}
\newcommand{\suptwo}[2]%
{\mathop{\sup_{#1}}_{#2}}
\newcommand{\supthree}[3]%
{\mathop{\mathop{\sup_{#1}}_{#2}}_{#3}}
\newcommand{\supfour}[4]%
{\mathop{\mathop{\mathop{\sup_{#1}}_{#2}}_{#3}}_{#4}} 
\newcommand{\inftwo}[2]%
{\mathop{\inf_{#1}}_{#2}}
\newcommand{\infthree}[3]%
{\mathop{\mathop{\inf_{#1}}_{#2}}_{#3}}
\newcommand{\inffour}[4]%
{\mathop{\mathop{\mathop{\inf_{#1}}_{#2}}_{#3}}_{#4}} 

\newcommand\calB{{\cal B}}

\newcommand\calE{{\cal E}}

\newcommand{\bsp}{\boldsymbol{p}}
\newcommand{\bsq}{\boldsymbol{q}}

\newcommand{\bbR}{\mathbb{R}}

\newcommand{\ep}{\varepsilon}

\newcommand{\bcs}{\backslash}

\newcommand{\para}[1]{{\bf #1}\/.---}
\newtheorem{thm}{Theorem}
\newtheorem{lm}[thm]{Lemma}
\newtheorem{pro}[thm]{Proposition}

\newcommand{\bthm}[1]{\begin{thm} #1 \end{thm}}
\newcommand{\blm}[1]{\begin{lm} #1 \end{lm}}
\newcommand{\bpro}[1]{\begin{pro} #1 \end{pro}}

\theoremstyle{definition}
\newtheorem{dfn}{Definition}
\newcommand{\bdf}[1]{\begin{dfn} #1 \end{dfn}}
\newtheorem{cjt}{Conjecture}
\newcommand{\bcjt}[1]{\begin{cjt} #1 \end{cjt}}

\newcommand{\rhoG}{\rho_{\rm Gibbs}}
\newcommand{\test}{t_{\rm est}}

\newcommand{\pG}{\bsp_{\rm Gibbs}}

\def\rnum#1{\resizebox{0.5em}{\height}{\expandafter{\romannumeral #1}}}
\def\Rnum#1{\resizebox{0.5em}{\height}{\uppercase\expandafter{\romannumeral #1}}}

\newcommand{\titlename}{Correlated catalyst in quantum thermodynamics}

\title{\titlename}

\author{Naoto Shiraishi}
\institute{Department of Basic Science, The University of Tokyo, 3-8-1 Komaba, Meguro-ku, Tokyo 153-8902, Japan}%

\date{\today}

\abstract{
In this short review article, we present recent progress in quantum thermodynamics in the framework with a correlated catalyst.
We examine two key properties of thermal operations, the Gibbs preserving property and the covariant property.
The state convertibility of a Gibbs preserving operation is fully characterized by the second law of thermodynamics with the nonequilibrium free energy.
The state convertibility of a covariant operation is shown to be free as long as an initial state has finite coherence.
We finally show that these two findings can be combined in the enhanced thermal operation (covariant Gibbs-preserving operation).
}

\begin{document}


\maketitle

In conventional macroscopic thermodynamics, the necessary and sufficient condition of state conversion is given by the second law of thermodynamics.
The entropy (adiabatic case) and the free energy (isothermal case) serve as the unique thermodynamic potential.
In quantum thermodynamics, we investigate the necessary and sufficient condition of state conversion and the robustness of the second law of thermodynamics in small quantum systems~\cite{Gour-review, Los-review, Sag-review}.
To respect thermodynamic aspects, in quantum thermodynamics, we can address only a limited class of quantum operations that embody thermodynamic processes.
In terms of the resource theory, a class of operations that we can address is called {\it free operations}.
The goal of quantum thermodynamics is to clarify whether a given state $\rho$ is convertible to another given state $\sigma$ through a thermodynamically free operation.

In quantum thermodynamics, we frequently employ the following two classes of free operations.
In the following, we consider a system $S$ with Hamiltonian $H$.

\bdf{[Gibbs-preserving operation (GPO)]
A CPTP map $\calE: S\to S$ is called a Gibbs-preserving operation (GPO) if $\calE(\rhoG)=\rhoG$.
Here, we defined the Gibbs state $\rhoG=e^{-\beta H}/Z$ for a given inverse temperature $\beta$.
}
\bdf{[Thermal operation (TO)]
A CPTP map $\calE: S\to S$ is called a thermal operation (TO) if there exists a proper auxiliary system $B$ with Hamiltonian $H_{\rm B}$ and an energy-conserving unitary $U$ on the composite system $SB$ such that $\calE(\rho)=\Tr_B[U(\rho\otimes \rhoG^{\rm B})U^\dagger]$.
Here, $\rhoG^{\rm B}$ is the Gibbs state of $B$.
}

These two operations are axiomatic (top-down) and operational (bottom-up) characterizations of thermodynamic processes.
It is easy to confirm that if an operation is TO, it is GPO.
The converse is also true if we consider the classical regime.
Here, we say that our subject is {\it classical} if both the input and the output states are diagonal with respect to an energy basis.

\bthm{[Horodecki-Oppenheim~\cite{HO13}, Shiraishi~\cite{Shi20}]
Consider conversions in the classical regime.
For any GPO $\calE$ and for any $\ep>0$, there exists TO $\calE'$ such that $\sup_\rho \abs{\calE(\rho)-\calE'(\rho)}_1<\ep$.
}

On the other hand, GPO and TO have a gap in the fully quantum regime~\cite{FOR15}.
The reason for this gap is considered to come from quantum coherence.
In fact, TO cannot convert an energy eigenstate $\ket{E_0}$ into a superposition of two energy eigenstates $\frac{1}{\sqrt{2}}(\ket{E_0}+\ket{E_1})$, while some GPO can.

This restriction has been investigated in the resource theory of (unspeakable) coherence, or $U(1)$ asymmetry~\cite{Jan06, GS08, Mar-thesis, MS14, MS16}.
In this resource theory, a free operation is a covariant operation, which can be implemented with an energy-conserving unitary and a diagonal state with an energy basis:

\bdf{[Covariant operation]
A CPTP map $\calE: S\to S$ is called a covariant operation if $\calE(e^{-iHt}\rho e^{iHt})=e^{-iHt}\calE(\rho) e^{iHt}$ for any $t$.
}
\bdf{[Incoherent state]
A state $\rho$ is incoherent if $e^{-iHt}\rho e^{iHt}=\rho$ for any $t$.
}

In coherence manipulation, it is well known that both axiomatic and operational characterizations coincide.
For its proof, see, e.g., Ref.~\cite{Mar-thesis}:

\bthm{
For any covariant operation $\calE$, there exists a proper auxiliary system $A$ with Hamiltonian $H_{A}$, an incoherent state $\xi$ on $A$, and an energy-conserving unitary $U$ on the composite system $SA$ such that $\calE(\rho)=\Tr_A[U(\rho\otimes \xi)U^\dagger]$.
}

Under the covariant condition, we cannot convert an energy eigenstate into a superposition of multiple energy eigenstates without any additional help, which serves as a severe restriction on possible state transformations.
In fact, the covariant property is considered to be one of the two major characterizations of quantum TO besides the Gibbs-preserving property~\cite{Los-review, Los15}.


It is known that in the small systems, various barriers other than the conventional second law emerge and limit our power of state conversion~\cite{Bra15, AN, Kli, Tur, MOAbook, FR18}.
On the other hand, an auxiliary system $C$ called {\it catalyst} may enhance state convertibility.
Here, the catalyst is a system which does not change its own state, while it helps the state conversion in the system.
In particular, in this article we investigate state convertibility with a {\it correlated catalyst}\cite{Mul18, SS21, Wil21, LS21, KDS21, LJ21, Cha21, Wil22, TS22, GKS23, ST23, KGS23} (see also review papers~\cite{Dat22, BWN23}), where we allow arbitrarily small correlation between the system and the catalyst in the final state.
A correlated catalyst removes various fragile constraints and raises only physically relevant constraints.

\bdf{[correlated catalyst]
We say that a class of free operations $X$ transforms $\rho$ to $\rho'$ with a correlated catalyst with vanishing error if for any $\ep>0$ and $\delta>0$ there exist a catalyst $C$, its state $c$, and a CPTP map $\calE: SC\to SC$ such that $\tau=\calE(\rho\otimes c)$ with $\Tr_S[\tau]=c$, $\abs{\Tr_C[\tau]-\rho'}_1<\ep$, and $\abs{\tau-\Tr_C[\tau]\otimes c}_1<\delta$.
}

In the above definition, we allow an arbitrarily small error in the system, which we call {\it vanishing error}.
In the case of {\it uncorrelated} catalyst, we replace the condition $\tau=\calE(\rho\otimes c)$ with $\Tr_S[\tau]=c$, $\abs{\Tr_C[\tau]-\rho'}_1<\ep$ by $\calE(\rho\otimes c)=\kappa\otimes c$ with $\abs{\kappa-\rho'}_1<\ep$.


In this article, we review the recent progress of quantum thermodynamics in the framework with a correlated catalyst, as explaining their proof outlines.
We first analyze the Gibbs preserving operation in both the classical and the quantum regime, and then explain the coherence manipulation by covariant operations.
We finally comment on the problem of quantum thermal operation, the ultimate goal.

\medskip

\para{Quantum thermodynamics in the classical regime}
We start with conversions with an uncorrelated catalyst in the classical regime.
A key breakthrough is the following result on a doubly-stochastic matrix  for exact conversion.


\bthm{[Klimesh~\cite{Kli}, Turgut~\cite{Tur}]\lb{t:trump}
Consider full rank distributions $\bsp$ and $\bsp'$ such that $\bsp'$ is not a permutation of $\bsp$.
Then, $\bsp$ is convertible to $\bsp'$ exactly by a doubly-stochastic matrix with an uncorrelated catalyst if and only if $f_\alpha(\bsp)>f_\alpha(\bsp')$ holds for all $\alpha\in \bbR$ with
\eq{
f_\alpha(\bsp):=
\bcases{
s_\alpha \ln \sum_i p_i^\alpha & \alpha \neq 0,1 \\
\sum_i p_i\ln p_i &\alpha=1 \\
-\sum_i \ln p_i &\alpha=0,
}
}
where $s_\alpha=1$ for $\alpha>1$ and $\alpha<0$, and $s_\alpha=-1$ for $0<\alpha<1$.
}

It is noteworthy that$f_\alpha$ for $\alpha\neq 0$ coincides with a slight modification of the well-known R\'{e}nyi $\alpha$-entropy $S_\alpha=\frac{1}{1-\alpha}\ln \sum_i p_i^\alpha $.
The quantity $f_0$ is known as Burg entropy~\cite{Burg}. 

The proof of \tref{trump} is highly complicated, and at present no simple proof is known.
The proof idea is to construct a catalyst satisfying a majorization relation (a condition for conversion without a catalyst).

We here introduce conversions of a pair of states.
We define the catalytic conversion of a pair of two states as follows.
(To keep quantum extension in mind, we here provide the definition for quantum states.)

\bdf{[catalytic conversion of a pair of states]
We say that $(\rho, \sigma)$ can be converted to $(\rho', \sigma')$ with a correlated catalyst with a vanishing error if for any $\ep>0$ and $\delta>0$ there exist a catalyst $C$, its state $c$ and $c'$ with ${\rm supp}(c)\subset {\rm supp}(c')$, and a CPTP map $\calE: SC\to SC$ such that $\tau=\calE(\rho\otimes c)$ with $\Tr_S[\tau]=c$, $\abs{\Tr_C[\tau]-\rho'}_1<\ep$, $\abs{\tau-\Tr_C[\tau]\otimes c}_1<\delta$, and $\calE(\sigma\otimes c')=\sigma'\otimes c'$.
}

In the case with an uncorrelated catalyst and exact conversion, we modify the condition on the transformation of $\rho$ with keeping that on $\sigma$.
In this definition, we keep in mind an application to GPO.
If both $\sigma$ and $\sigma'$ are the Gibbs state of $S$, then $\calE$ in the above definition is a GPO by setting the Hamiltonian of the catalyst $C$ such that $c'$ is a Gibbs state of $C$.
This answers why we do not allow correlation and error in the transformation of $\sigma$.

In the conversion of a pair of two classical states $(\bsp, \bsq)$ to $(\bsp', \bsq')$, if all the entries of $\bsq$ and $\bsq'$ are rational numbers, we have a useful technique called {\it trivialization}.
Suppose $q_i=n_i/N$ and $q'_i=n'_i/N$.
Then, by regarding that the state space has $N$ small states and grouping $n_i$ small states as state $i$ in $S$, we can reduce to the case of doubly-stochastic matrices.
With accepting vanishing error in the final state, \tref{trump} leads to the following result:

\bthm{[Brand\~{a}o {\it et al}.~\cite{Bra15}]\lb{t:second-laws}
Consider full rank distributions $\bsp$, $\bsq$, $\bsp'$, and $\bsq'$.
Then, $(\bsp, \bsq)$ is convertible to $(\bsp', \bsq')$ with vanishing error with an uncorrelated catalyst if and only if
\eq{
S_\alpha(\bsp||\bsq)\geq S_\alpha(\bsp'||\bsq')
}
holds for all $-\infty< \alpha<\infty$, where $S_\alpha(\bsp ||\bsq):={\rm sgn}(\alpha)/(\alpha-1)\ln (\sum_i p_i^\alpha/q_i^{\alpha-1})$ ($\alpha\neq 0,1,\pm\infty$) is the R\'{e}nyi $\alpha$-divergence.
The cases with $\alpha= 0,1,\pm\infty$ are defined by their limits.
}

By setting $\bsq=\pG$, $S_\alpha(\bsp ||\bsq)$ reads an extended free energy $F_\alpha(\bsp)$, since its $\alpha=1$ case is equivalent to the nonequilibrium free energy besides constant.
We usually express $F_1(\bsp)$ simply by $F(\bsp)$.
\tref{second-laws} indicates that an infinite family of second laws serves as the necessary and sufficient condition for state conversions through GPO.


We now consider the case with a correlated catalyst.
Importantly, with correlation most of R\'{e}nyi $\alpha$-divergence violates subadditivity.
Let $\bsp^{\rm AB}$ be a probability distribution on a composite system $AB$ whose marginal distributions on $A$ and $B$ are $\bsp^{A}$ and $\bsp^{\rm B}$ respectively.
Then, the subadditivity of the R\'{e}nyi $\alpha$-divergence
\eq{
S_\alpha(\bsp^{AB}||\bsq^A\otimes \bsq^B)\geq S_\alpha(\bsp^{A}||\bsq^A)+S_\alpha(\bsp^{B}||\bsq^B),
}
or equivalently $F_\alpha(\bsp^{AB})\geq F_\alpha(\bsp^{A})+F_\alpha(\bsp^{B})$, is satisfied only with $\alpha=0$ and 1, and is violated for all other $\alpha$.
In addition to this, we notice that $F_0$ is fragile against a small perturbation.
In other words, only the constraint with $\alpha=1$ is robust, and all other constraints are fragile against negligibly small correlation or perturbation.
The above implication is indeed correct, and we recover the second law of thermodynamics where a single second law inequality determines the state convertibility.

\bthm{[M\"{u}ller~\cite{Mul18}]\lb{t:c-second-law}
A probability distribution $\bsp$ is convertible to $\bsp'$ through a GPO with a correlated catalyst if and only if $F(\bsp)\geq F(\bsp')$.
}

This comes from a more general result on exact conversions, and \tref{c-second-law} can be shown by applying trivialization.

\bthm{[M\"{u}ller~\cite{Mul18}]\lb{t:c-exact}
Consider full rank distributions $\bsp$ and $\bsp'$ such that $\bsp'$ is not a permutation of $\bsp$.
Then, $\bsp$ is convertible to $\bsp'$ exactly by a using doubly-stochastic matrix with a correlated catalyst if and only if $H_0(\bsp)\leq H_0(\bsp')$ and $S_1(\bsp)<S_1(\bsp')$ hold.
Here, $H_0(\bsp):=\ln \#\{i|p_i\neq 0\}$ is the {\it max entropy}.
}

\tref{c-second-law} establishes the existence of a single thermodynamic potential as in the macroscopic thermodynamics.
Only the constraint with the conventional nonequilibrium free energy is relevant, and all the other fragile effects are removed by the setting with a correlated catalyst.

The original proof of Theorem \ref{t:c-second-law} shown in \cite{Mul18} is highly complicated such that we construct an elaborated catalyst satisfying all the conditions in \tref{trump}.
Later, another simple proof of Theorem~\ref{t:c-second-law} is presented in \cite{SS21}, which applies to both classical and quantum setups.
We shall see the idea of its proof in the next section.

\medskip

\para{Gibbs-preserving operation in quantum regime}
Unlike the classical regime, a simple criterion on state convertibility by GPO is not known in the quantum regime.
A big stumbling block is the absence of a Lorentz curve for quantum density matrices, which prevents a clear characterization of state convertibility.
Due to this deficit, we cannot reach a result with a correlated catalyst by simply quantizing the approach for the classical case.

In spite of this trouble, it was conjectured that the same condition as the classical regime (\tref{c-second-law}) is still valid in the fully quantum regime~\cite{WGE17, LM19}.
This conjecture is solved in positive in \cite{SS21}:

\bthm{[Shiraishi-Sagawa~\cite{SS21}]\lb{t:q-second-law}
A quantum state $\rho$ is convertible to $\rho'$ by a GPO with a correlated catalyst if and only if $F(\rho)\geq F(\rho')$.
}

Here, $F(\rho):=\frac{1}{\beta}(S_1(\rho||\rhoG)-\ln Z)$ is the nonequilibrium free energy at inverse temperature $\beta$ with quantum relative entropy $S_1(\rho||\rhoG):=\Tr[\rho \ln \rho- \rho \ln \rhoG]$, where $Z$ is a partition function.
This theorem establishes the recovery of the second law of thermodynamics in the small quantum regime.
This theorem is obtained as a special case of the following theorem:

\bthm{[Shiraishi-Sagawa~\cite{SS21}]\lb{t:q-pair}
Consider states $\rho$, $\sigma$, $\rho'$, and $\sigma'$ with ${\rm supp}(\rho)\subset {\rm supp}(\sigma)$.
Then, $(\rho, \sigma)$ is convertible to $(\rho', \sigma')$ with vanishing error with a correlated catalyst if and only if $S_1(\rho||\sigma)\geq S_1(\rho'||\sigma')$.
}

As mentioned above, this theorem is proven by a completely different approach from that presented in the previous section.
In this proof, we directly treat conversions with a correlated catalyst without dealing with conversions with an uncorrelated catalyst or without a catalyst.
Below we outline the proof of this theorem, which also serves as a basic proof technique for other resource theories with a correlated catalyst.


We here only prove the {\it if} part, which is the difficult part of this proof.
To this end, we first show the convertibility in the asymptotic conversion, and then reduce this result to the catalytic conversion.

We start with a sufficient (but not necessary) condition of state conversions of a pair of states.

\bdf{[quantum R\'{e}nyi $0/\infty$-divergence]
The quantum R\'{e}nyi $0/\infty$-divergence are defined as
\balign{
S_0(\rho||\sigma)&:=-\ln \Tr[P_\rho \sigma], \\
S_\infty(\rho||\sigma)&:=\ln [\min\{ \lmd | \rho\leq \lmd \sigma\}],
}
where $P_\rho$ is a projector on the support of $\rho$.
}

\blm{[Faist-Renner~\cite{FR18}]\lb{t:suffice-conv}
There exists a CPTP map $\Lambda$ satisfying $\rho'=\Lambda(\rho)$ and $\sigma'=\Lambda(\sigma)$ if
\eqa{
S_0(\rho||\sigma)\geq S_\infty(\rho||\sigma).
}{suffice-conv-cond}
}

Below we present an intuitive reason why \lref{suffice-conv} holds.
A state conversion is known to always decrease state distinguishability.
The R\'{e}nyi $\alpha$-divergence serves as a measure of distinguishability of two states, and a larger $\alpha$ gives a larger value.
Although the quantification of distinguishability has some range, we know that the smallest is $\alpha=0$ and the largest is $\alpha=\infty$.
Thus, \eref{suffice-conv-cond} implies that $\rho$ and $\sigma$ are easier to distinguish in any sense than $\rho'$ and $\sigma'$.
From this observation, conversion $(\rho,\sigma)$ to $(\rho', \sigma')$ does not conflict with the decrease of distinguishability, which supports the presence of a conversion protocol, and this indeed exists.

Our next tool is the convergence of the $\ep$-smoothed R\'{e}nyi $\alpha$-divergence rate to the quantum relative entropy, which stems from the quantum Stein's lemma.
Let $\calB_\ep(\rho):=\{ \kappa| \abs{\kappa-\rho}\leq \ep\}$ be a ball with diameter $\ep$ with the center at $\rho$.
Then we define the $\ep$-smoothed R\'{e}nyi $0/\infty$-divergence as $S_0^\ep (\rho||\sigma):=\max_{\kappa\in \calB_\ep(\rho)} S_0 (\kappa||\sigma)$ and $S_\infty^\ep(\rho||\sigma):=\min_{\kappa\in \calB_\ep(\rho)} S_\infty (\kappa||\sigma)$.
The smoothing process plays the role of removing some singularity in R\'{e}nyi divergences.


\blm{[Nagaoka-Hayashi~\cite{NH07}, Datta~\cite{Dat09}]\lb{t:Renyi-converge}
For any $0<\ep<1/2$, $\ep$-smoothed R\'{e}nyi $0/\infty$-divergence rate converges to the relative entropy:
\eq{
\lim_{n\to \infty} \frac1n S_0^\ep (\rho^{\otimes n}||\sigma^{\otimes n})=\lim_{n\to \infty} \frac1n S_\infty^\ep (\rho^{\otimes n}||\sigma^{\otimes n})=S_1(\rho||\sigma).
}
}

This lemma is strongly related to quantum hypothesis testing.
The task of quantum hypothesis testing is to distinguish two quantum states, whose success probability is characterized by {\it hypothesis testing divergence}.
Roughly speaking, for small $\ep>0$, $\ep$-smoothed R\'{e}nyi $0$-divergence (resp. R\'{e}nyi $\infty$-divergence) is connected to the hypothesis testing divergence with accuracy close to $1$ (resp. 0) (see \cite{Sag-review}).
The quantum Stein's lemma~\cite{HP, ON} states that all the hypothesis testing divergence rates for accuracy $0<\eta<1$ converges to the relative entropy.
Combining this fact, we obtain \lref{Renyi-converge}.

Using these two lemmas, we find the necessary and sufficient condition for asymptotic conversions.
Suppose $S_1(\rho||\sigma)\geq S_1(\rho'||\sigma')$.
Owing to \lref{Renyi-converge}, there exists $n$ such that $S_0^\ep (\rho^{\otimes n}||\sigma^{\otimes n})\geq S_\infty^\ep (\rho'^{\otimes n}||\sigma'^{\otimes n})$.
Applying \lref{suffice-conv}, we conclude that there exists a CPTP map which converts $\sigma^{\otimes n}$ to $\sigma'^{\otimes n}$ and $\rho^{\otimes n}$ to close to $\rho'^{\otimes n}$.

\blm{\lb{t:GPO-asymptotic}
Suppose $S_1(\rho||\sigma)\geq S_1(\rho'||\sigma')$.
Then, for any $\ep>0$ there exists a sufficiently large $n$ and a CPTP map $\Lambda$ such that $\Lambda(\sigma^{\otimes n})=\sigma'^{\otimes n}$ and $\Lambda (\rho^{\otimes n})=\Xi$ with $\abs{\Xi-\rho'^{\otimes n}}_1<\ep$.
}

Our final task is to reduce the result of approximate asymptotic conversions to correlated-catalytic conversions.
To do this, we introduce a useful tool, which was first demonstrated in quantum thermodynamics~\cite{SS21}, and then applied to various setups including entanglement, teleportation, and entropy conjecture~\cite{Wil21, LS21, KDS21, LJ21, Cha21, Wil22, TS22}.

\blm{[Shiraishi-Sagawa~\cite{SS21}]\lb{t:asymptotic-catalytic}
Suppose that there exists a CPTP map $\Lambda: S^{\otimes n}\to S^{\otimes n}$ such that $\Lambda(\sigma^{\otimes n})=\sigma'^{\otimes n}$ and $\Lambda (\rho^{\otimes n})=\Xi$ with $\abs{\Xi-\rho'^{\otimes n}}_1<\ep$.
Then, $(\rho,\sigma)$ is convertible to $(\rho', \sigma')$ through a CPTP map with a correlated catalyst with vanishing error.
}

This lemma is proven by constructing the desired catalyst explicitly.
Noting $\Xi$ is a state on $S_1\otimes S_2\otimes\cdots \otimes S_n$, we introduce $\Xi_i$ ($1\leq i\leq n$) as a reduced state of $\Xi$ on $S_1\otimes \cdots \otimes S_i$.
Introducing a label system $R$ whose Hamiltonian is trivial, the catalyst $c$ on $S^{\otimes n-1}\otimes R$ is constructed as
\eq{
c:=\frac1n \sum_{k=1}^n \rho^{\otimes k-1}\otimes \Xi_{n-k}\otimes \ket{k}\bra{k}.
}
where $\ket{k}$ is a state on $R$.
Then, the initial state of $SC$ reads $\rho\otimes c=\frac1n \sum_{k=1}^n \rho^{\otimes k}\otimes \Xi_{n-k}\otimes \ket{k}\bra{k}$.
By applying $\Lambda$ if the label is $\ket{n}$, the state becomes $\tau'=\frac1n \sum_{k=1}^n \rho^{\otimes k-1}\otimes \Xi_{n+1-k}\otimes \ket{k}\bra{k}$.
A proper relabeling of the label system and $n$ copies of states leads to \lref{asymptotic-catalytic}.
This completes the proof of \tref{q-pair}.

The connection between exact asymptotic conversions and uncorrelated-catalytic conversions was first discussed by Duan {\it et al.}\cite{Dua}.
We note that even if the condition $\abs{\Xi-\rho'^{\otimes n}}_1<\ep$ for asymptotic conversions can be weakened as $\abs{\Tr_{\bcs i}[\Xi]-\rho'}_1<\ep$ for all $i$, which is called {\it marginal asymptotic conversions}~\cite{Fer23}, the same conversion protocol works and we obtain correlated-catalytic conversions.
This point was first explicitly pointed in Ref.~\cite{GKS23}.

\medskip

In addition to these, we further introduce a useful tool transforming an approximate conversion with vanishing error into an exact conversion, which was first argued by Wilming~\cite{Wil22} (see also \cite{TS22} for a general expression):

\blm{[Wilming~\cite{Wil22}]\lb{t:approximate-exact}
Consider a sequence of convex sets of quantum states $\{S_m\}_{m=1}^\infty$ satisfying $S_m\subseteq S_{m+1}$.
Let $V$ be the interior set of $\lim_{m\to \infty} S_m$.
If $\kappa$ is an interior state of $V$, then there exists an integer $m$ such that $\kappa\in S_m$.
}

Below we demonstrate how to use this formal theorem in the resource theory.
We set $S_m$ as a set of states convertible from $\rho$ exactly through a CPTP map with a correlated catalyst whose Hilbert space dimension is $m$ under the condition that $\sigma$ is convertible to $\sigma'$.
We have already found that $V=(\lim_{m\to \infty}S_m)^{\rm c} =\{ \rho'| F(\rho)\geq F(\rho')\}$, where superscript $c$ means its closure.
The boundary of $V$ consists of (i) pure states, and (ii) a set of states $\rho'$ with $S_1(\rho||\sigma)=S_1(\rho'||\sigma')$.
Hence, a full-rank $\rho'$ with $S_1(\rho||\sigma)>S_1(\rho'||\sigma')$ is an interior point of $V$, and thus we have $\rho'\in S_m$ for some $m$, meaning that $\rho'$ is obtained from $\rho$ with a correlated-catalyst with dimension $m$ exactly.
This leads to an exact conversion of a pair of states, which is a new result presented in this article.

\bthm{
Consider states $\rho$, $\sigma$, $\rho'$, and $\sigma'$ such that ${\rm supp}(\rho)\subset {\rm supp}(\sigma)$ and $\rho'$ is full rank.
If $S_1(\rho||\sigma)> S_1(\rho'||\sigma')$, then $(\rho, \sigma)$ is convertible to $(\rho', \sigma')$ exactly with a correlated catalyst.
}

In terms of quantum thermodynamics, our claim is that we can convert $\rho$ to $\rho'$ by a GPO exactly if $F(\rho)>F(\rho')$ and $\rho'$ is full rank.
We note that the above condition that $\rho'$ is full rank cannot be replaced by $H_0(\rho)\leq H_0(\rho')$, as \tref{c-exact}.
A simple counterexample is a case that both $\rho$ and $\rho'$ are pure states, where a pure state $\rho'$ cannot correlate with a catalyst and thus additional constraints in the case of an uncorrelated catalyst recover.

\medskip

\para{Coherence under covariant operation}
We next investigate state conversions through a covariant operation.
From the definition, a covariant operation without any help cannot increase the amount of coherence among energy eigenstates~\cite{Mar-thesis}.
This implies that coherence among energy eigenstates is a precious resource that is not obtained freely.
Note that the law of energy conservation is ubiquitous, and thus we face the above problem in all natural setups.

We shall see how this severe restriction on coherence transformation is changed or not with the help of a catalyst.
We first see some results where a correlated catalyst puts no additional power, which implies the hardness of coherence manipulation.

\bthm{[No broadcasting theorem (Lostaglio-M\"{u}ller~\cite{LM19}, Marvian-Spekkens~\cite{MS19})]
Suppose that $\rho$ is incoherent and $\rho$ is converted to $\rho'$ through a covariant operation with a correlated catalyst.
Then, $\rho'$ is also incoherent.
}

The proof of no broadcasting theorem is not long but highly technical.
We apply the Koashi-Imoto decomposition~\cite{KI02} to the composite system $SC$ and show that this conversion does not touch the coherence in the system.

No broadcasting theorem is applicable only when the system is completely incoherent.
However, it is natural to expect that if two modes (energy level spacing) are relatively irrational (e.g., mode 1 and mode $\sqrt{2}$), then coherence on one mode (e.g., mode 1) does not help to create coherence on another mode (e.g., mode $\sqrt{2}$).
This expectation is indeed correct, and we have the following {\it no mode-broadcasting theorem}.

\bthm{[No mode-broadcasting theorem (Shiraishi-Takagi~\cite{ST23})]
Suppose that $\rho'$ has coherence on mode $a$ and $\rho$ has no coherence on modes which are rational multiples of $a$.
Then, $\rho$ is not convertible to $\rho'$ through a covariant operation with a correlated catalyst.
}

No mode-broadcasting theorem is proven by contradiction.
We suppose contrarily that a protocol violating no mode-broadcasting theorem exists, and using this protocol we construct a protocol violating no broadcasting theorem.

As seen above, a correlated catalyst provides no additional help if the initial state is incoherent.
Then, a natural question is what happens if an initial state $\rho$ has some coherence.
Surprisingly, in this case, we have no obstacle to state conversion, and a state with negligibly small coherence can be converted into a maximally coherent state through a covariant operation with a correlated catalyst.
The following theorem clearly shows a striking fact that the only meaningful distinction in the resource theory of coherence is whether the system has finite coherence or indeed no coherence, and if the system has finite coherence, its amount is irrelevant.

\bthm{[Shiraishi-Takagi~\cite{ST23}, Kondra-Ganardi-Streltsov~\cite{KGS23}]\lb{t:coherence-trivial}
Suppose that all coherent modes in $\rho'$ are integer multiples of coherent modes in $\rho$.
Then, $\rho$ is convertible to $\rho'$ with vanishing error through a covariant operation with a correlated catalyst.
}

In addition to this, \cite{ST23} also shows that if $\rho'$ is full rank, then this conversion is exact.

Below we outline the proof of this theorem.
Our starting point is the coherence amplifier in a two-level system, which serves as a subroutine in our protocol.

\blm{[Ding-Hu-Fang,~\cite{DHF21}]\lb{t:amplifier}
Let $\rho(a)=\mx{1/2&a/2\\ a/2&1/2}$ be a state of a two-level system in the energy basis $\{ \ket{0}, \ket{1}\}$.
For $0<a<1$, there exists a correlated-catalytic conversion from $\rho(a)$ to $\rho(a')$ with $a'=(25a-a^3)/24>a$.
}

Using this amplifier repeatedly, we can realize arbitrary state transformation with {\it marginal catalysts} by using only a covariant operation.
Marginal catalysts are multiple catalysts $c_1,c_2,\ldots$ which can correlate with the system and other catalysts at the final state but return to the original state by taking its reduced state.
The idea of marginal catalysts was first introduced in the context of quantum thermodynamics in the classical regime~\cite{LMP15,MP}.
A successive application of the amplifier on these systems produces many almost maximally coherent states arbitrarily close to $\frac{1}{\sqrt{2}}(\ket{0}+\ket{1})$.

Since we can distribute coherence in a single state to many incoherent states and make them coherent, we can prepare Gaussian-like state $[\frac{1}{\sqrt{2}}(\ket{0}+\ket{1})]^{\otimes n}$ with sufficiently large $n$ as an external system from a single coherent state.
Employing this Gaussian-like state to absorb the backaction of energy, we can emulate any unitary channel as shown in \cite{AS67, KMP04, MM08, Abe14, TSS20}, which enables us to prepare any state from an incoherent state.
Noting that two marginal catalysts create a coherent state from scratch, we arrive at the following result:

\bthm{[Takagi-Shiraishi~\cite{TS22}]\lb{t:marginal-catalyst}
For any $\rho$ and $\rho'$, a covariant operation converts $\rho$ to $\rho'$ with vanishing error with marginal catalysts.
}

Next, we show that there exists a marginal asymptotic conversion protocol which transforms any coherent state $\rho$ to any state $\rho'$.
To construct this protocol, we first distill a set of the marginal catalysts $c_1, c_2,\ldots , c_M$ used in \tref{marginal-catalyst} from (maybe many but) finite copies of $\rho$, with which we can convert an incoherent state $\zeta$ to the desired state $\rho'$.
This distillation succeeds if $\rho$ has finite coherence on proper modes, and with the help of \lref{approximate-exact} this distillation can be exact.
Then, if we have $M^k$ sets of marginal catalysts, reusing these catalysts with recombination we can perform the conversion $\zeta\to \rho'$ not $M^k$ but $kM^k$ times.
By setting $k$ sufficiently large, the transformation rate from $\rho$ to $\rho'$ can be arbitrarily large, and in particular, a marginal asymptotic conversion protocol with transformation rate 1 exists.
A similar idea for the case with a correlated catalyst is seen in \cite{GKS23}, and the above protocol is its extension.

Finally, applying \lref{asymptotic-catalytic} to this marginal asymptotic conversion, we obtain the desired correlated-catalytic conversion protocol.

\medskip

We note that if all the energy level spacings in the system are integer multiples of a fixed value, we have another protocol of a marginal-asymptotic conversion from coherent $\rho$ to any $\rho'$ with any transformation rate.
A key fact is that an optimal phase estimation protocol for $\rho^{\otimes n}$ has variance $O(1/n)$.

\bpro{\lb{t:cov}
Suppose that all the energy level spacings of system $S$ are integer multiples of a constant $\Delta$.
Then, there exists a time estimation protocol which outputs a real number $\test$ with probability distribution $P(\test|\kappa)$ for $\kappa$ on $S^{\otimes n}$ such that (i) $P(\test|e^{-iH^{\otimes n}\tau}\kappa e^{iH^{\otimes n}\tau})=P(\test+\tau|\kappa)$ for any $\kappa$ and $\tau$, and (ii) the variance of $P(\test|\rho^{\otimes n})$ decays as $O(1/n)$ for any $\rho$ with coherence on mode $\Delta$.
}

Now we shall construct a sublinear asymptotic conversion protocol $\rho^{\otimes n}\to \rho'^{\otimes m}$ with $\lim_{n\to \infty}\frac mn=0$ as follows.
We first estimate $\test$ by \pref{cov} with $\rho^{\otimes n}$, and then prepare $e^{-iH^{\otimes m}\test}(\rho')^{\otimes m}e^{iH^{\otimes m}\test}$.
This protocol is covariant by construction, and the output state is close to $(\rho')^{\otimes m}$ with arbitrarily small error:

\bthm{[Marvian~\cite{Mar20}]
Suppose that all the energy level spacings of system $S$ are integer multiples of a constant $\Delta$.
Then, for any $\rho$ with coherence on mode $\Delta$, any $\rho'$, any function $m(n)$ with $\lim_{n\to \infty}\frac{m(n)}{n}=0$ and any $\ep>0$, there exists a sufficiently large $n$ and a covariant operation $\Lambda$ such that $\abs{\Lambda(\rho^{\otimes n})-\rho'^{\otimes m(n)}}_1<\ep$.
}

To construct a marginal-asymptotic conversion protocol with rate $r$, we first estimate $\test$ from $\rho^{\otimes n}$ and prepare $\frac{rn}{m(n)}$ copies of $e^{-iH^{\otimes m}\test}(\rho')^{\otimes m}e^{iH^{\otimes m}\test}$ with using the same $\test$.
In marginal-asymptotic conversions, the correlation among copies is irrelevant.

\medskip

\para{Toward quantum thermal operation}
We have seen how two characterizations of quantum thermal operation, the Gibbs-preserving property and the covariant property, restrict state convertibility.
\tref{q-second-law} states that the Gibbs-preserving property provides the second law of thermodynamics with the nonequilibrium free energy.
\tref{coherence-trivial} states that coherence provides no restriction as long as an initial state has finite coherence.

A class of operations satisfying both the Gibbs-preserving property and the covariant property is called {\it enhanced thermal operation} (EnTO)~\cite{Cwi15}, which is a slightly larger class than the original TO.
Owing to its axiomatic characterization, EnTO is also studied as an alternative to TO~\cite{Cwi15, Los15, Gou18, WT24}.
The convertibility condition for EnTO without catalyst has been derived in \cite{Gou18}, while it is highly complicated and not easy to handle as the case of quantum GPO.
Interestingly, the convertibility condition for EnTO with a correlated catalyst is given by the combination of that for GPO (\tref{q-second-law}) and covariant operations (\tref{coherence-trivial}):

\bthm{[Shiraishi~\cite{Shi24}]
Suppose that all the energy level spacings of system $S$ is integer multiples of a constant $\Delta$, and the initial state $\rho$ has coherence on mode $\Delta$.
Then, $\rho$ is convertible to $\rho'$ through an EnTO with a correlated catalyst if and only if $F(\rho)\geq F(\rho')$.
}

We outline the construction of this protocol.
From \pref{asymptotic-catalytic}, it suffices to show a marginal-asymptotic conversion protocol from $N=ma+d$ copies of $\rho$ with $a\ll d\ll ma$ to $ma$ copies of $\rho'$.
We set $a$ sufficiently large such that \lref{GPO-asymptotic} holds for $n=a$.
We denote this GPO by $\Lambda$.
Using two $\rho^{\otimes d/2}$, we estimate the phase and obtain estimators $\test^1$ and $\test^2$.
We then prepare $m$ copies of a channel $e^{iH^{\otimes a}\test^2}\Lambda(e^{-iH^{\otimes a}\test^1}\kappa e^{iH^{\otimes a}\test^1})e^{-iH^{\otimes a}\test^2}$ for $\kappa$ on $S^{\otimes a}$.
It is easy to confirm that this channel is covariant, Gibbs-preserving, and maps $\rho^{\otimes (ma+d)}$ to $\rho'^{\otimes ma}$ in the marginal asymptotic sense.

A stronger statement for the original TO was conjectured as follows.

\bcjt{[Kondra-Ganardi-Streltsov~\cite{KGS23}]
$\rho$ is convertible to $\rho'$ through a TO with a correlated catalyst if and only if (i) $F(\rho)\geq F(\rho')$ and (ii) all coherent modes in $\rho'$ is integer multiples of coherent modes in $\rho$.
}

There are both positive and negative implications to this conjecture.
An opposing observation is that TO and EnTO have different state convertibility without catalyst~\cite{Cwi15, DDH21}, whose gap cannot be filled even in the case of a small error.
In spite of this negative suggestion, this conjecture may still hold since a correlated catalyst removes many fragile gaps and makes various different resource theories identical.
This gap closure indeed happens in some subclasses of TO~\cite{SN23}.
Therefore, even though To and EnTO accompany different resource theories without catalysts, it is still possible to expect that they collapse into the same theories with a correlated catalyst.
Another optimistic view is provided from results on asymptotic conversions, where convertibility by TO with some assists~\cite{Fai19, Sag21} or in some restricted setup~\cite{Gou22} is shown to be characterized by free energy.
Whether convertibility by TO with a correlated catalyst is also characterized by the free energy is left as a challenging open problem.

\end{document}